\documentstyle[12pt,dina4p]{article}
\begin{document}
\newcommand{\gl}[1]{(\ref{#1})}
\newcommand{\num}{\#}
\newcommand{\be}{\begin{equation}}
\newcommand{\ee}{\end{equation}}
\newcommand{\bea}{\begin{eqnarray}}
\newcommand{\eea}{\end{eqnarray}}
\newcommand{\eq}[1]{eq.~(\ref{#1})}
\newcommand{\fig}[1]{fig.~(\ref{#1})}
\newcommand{\tab}[1]{table~\ref{#1}}
\begin{centering}
  {\bf       Yet another way to obtain}\\
  {\bf  low temperature expansions for discrete spin systems}\\*[1cm]
           C. Vohwinkel\\*[1cm]
 Deutsches Elektronen-Synchrotron DESY, Hamburg\\*[2cm]
\end{centering}

\begin{abstract} \normalsize
I present a modification of the shadow-lattice technique, which
allows one to derive low temperature series for discrete spin models
to high orders. Results are given
for the 3-d Ising model up to 64 excited bonds, for the 4-d Ising model
up to 96 excited bonds and the 3-d Potts model up to 56 excited bonds.
\end{abstract} \normalsize

\section{}
Series expansions are a major tool to obtain critical properties of statistical
systems and to test Monte Carlo programs.
Usually series are derived using graphical techniques,
but recently a method using direct enumeration of the relevant
configurations on the lattice has been proposed~\cite{cre}.
Whereas the diagrammatic
techniques have the disadvantage that they are hard to port to parallel or
vector machines,
the direct enumeration relies heavily on fast machines,
taking full advantage of parallel architecture, but is computationally too
demanding to be implemented on smaller machines.
One advantage of direct enumeration is that it is done automatically,
i.e. once the program is set up it can in principle calculate the series
to any given order.

In this paper I shall sketch a method which is a variant of the
shadow method (see for example~\cite{domb})
 and which seems to be powerful, especially for higher dimensional
systems. A detailed description will be published elsewhere~\cite{voh}.
Although it is a graphical technique, the generation of the series
is fully automatic, minimizing possible errors due to missing diagrams.
Using this method, existing series for the Ising and
Potts model \cite{bha}\cite{pew}\cite{gut} are extended by more than ten terms
using 1 day of CPU-time on a personal computer.
To obtain the magnetisation and free energy for the $d=3$ Ising model up to
48 excited bonds I need 43 seconds, which has to be compared to more
than a day on a Cray-YMP, which was needed in~\cite{bha} to obtain the
free energy to the same order.
Because the method uses an embedding which
corresponds to real configurations, the calculation of correlation functions
could be added rather easily and, based on the experience of~\cite{pew}, I
expect an additional factor of 15 for the CPU-time needed.
The method could be reasonably well
adapted to vector computers, allowing the generation
of series for various discrete models with a small amount of CPU-time.

In the next section I shall describe the proposed method using the Ising model
as an example. In section~\ref{potts} the generalization to other
discrete models is addressed. Finally some results are presented.

\section{}
Consider the Ising model on a $d$-dimensional
hypercubic lattice
of volume $N$ in the presence of an external
magnetic field $h$. The spin variables on each site $\sigma(x)$
can take on the values $\pm 1$ and the action of the model
is given by
\vspace{1.0ex}
\be \label{eq201}
 S_{h} = -2\kappa \sum_{<xy>} \sigma(x) \sigma(y) - h\sum_{x} \sigma(x),
\ee
where the first sum runs over nearest
neighbours.

The free energy per unit volume $f$ is given by
\be \label{eq206}
 f(h) = - \lim_{N \rightarrow \infty} \frac{1}{N} \ln Z_{h}.
\ee

At couplings $\kappa\gg \kappa_c$ most of
the spins are aligned even with no magnetic
field. Thus one can start with a lattice of $N$ aligned spins and treat
excitations arising from groups of flipped spins.
For zero magnetic field
all states are doubly degenerate due
to the symmetry $\sigma \rightarrow -\sigma$. One can overcome this degeneracy
by introducing a small magnetic field.
With
$   u = e^{-8\kappa}  $ and $y=e^{-2h}$ the partition function for $N$ spins
can be rewritten as
\be \ln Z_N = -\frac{N}{2}\ln y -
\frac{qN}{8}\ln u + \sum_{r>0} P_{r,{\scriptscriptstyle N}}(y)u^r,\ee
where the $P_{r,{\scriptscriptstyle N}}(y)u^r$
are contributions from flipped
spins.

The task of the low temperature expansion
is to obtain the $P_{r,{\scriptscriptstyle N}}(y)$, giving an expansion
of thermodynamic quantities in $u$.
Because the free energy is an extensive quantity it suffices to calculate
\be \tilde{f}= \sum_{r>0} p_r(y)u^r,\ee
where $p_r(y)$ is the part of $P_{r,N}(y)$ linear in $N$.

One can divide the lattice into an even and an odd sub-lattice,
depending whether the sum of coordinates for a given site is even or odd.
The shadow method uses the fact that spins on odd sites do not interact
with each other:
If one flips a spin on an odd site, the change in the power of $u$ is
given by the number of its flipped neighbours, which are entirely on even
sites. It is therefore sufficient to embed
sub-configurations consisting only of spins on the even sub-lattice
and note the number $\omega(o)$
of flipped neighbours for each odd site $o$.
{}From now on I shall refer to odd sites as ``shadows'' and use the term
``spin'' exclusively for even sites.
We can break up the free energy into contributions $f_{s,t}$ coming from
$s$ flipped spins and $t$ flipped shadows:
\be \tilde{f}=\sum_{s,t} f_{s,t}.\ee
Because the even and the odd sub-lattice are symmetric, one must get
the same result when interchanging flipped spins and flipped shadows:
\be f_{s,t} = f_{t,s}. \ee
This property is referred to as code-balance.
The advantage of the shadow method is that one has to embed
approximately only half the spins with
little overhead for the bookkeeping of the $\omega(o)$.
In practice one saves less CPU-time than expected, because usually
all configurations up to a given number $s$ of spins are generated. The
resulting series are correct up to the lowest order one obtains with
$s+1$ spins. Most of the $s$-spin contributions have a degree higher
than that and are thus generated in vain. To give an example: only
1 out of 113636 (or even more) 8-spin diagrams is needed to calculate
the $d=4$ Ising model series up to order $32$.

Here I propose to use code-balance in conjunction with predetermining
the lowest degree of a diagram in order to generate only those diagrams
which are actually needed for a given maximum order $l_{max}$ in $u$.
Because code balance is employed one needs to find all diagrams $M$
which will give
non-zero contributions $f_{s,t}(M)$ with $s \leq t$ and an order in $u$
less than or equal to $l_{max}$.
Contributions $f_{s,t}$ with $s > t$ are given by code-balance.

Let me briefly review the steps necessary to construct a series:
A configuration on a lattice will consist of one or more clusters of flipped
spins where the different clusters are not joined by nearest-neighbour bonds,
but the spins within a cluster are connected among themselves.
The configurations are described by connectivity matrices. Because the shadow
method is used, only the spins on even sites are entered into the connectivity
matrix. An entry in the connectivity matrix describes whether two spins have
0, 1 or 2 shadows in common. Let me call a matrix describing $s$ spins an
$s$-matrix.
It is impractical to embed a configuration consisting of more than one cluster
because there will be of the order $N^c$ different relative positions
for $c$ clusters. Instead one constructs only connectivity matrices which
correspond to a single cluster and determines in a second step the number
of ways two or more of such matrices can be put on the lattice.

The construction of
the connectivity matrices can be done recursively: One starts with
the (trivial) matrix for one spin and adds another spin, which
must be connected
to the first spin by a non-zero entry in the connectivity matrix. Other spins
are then added in such a way that each new spin is connected to at least one
spin of the old matrix.

Having constructed a connectivity matrix, one can embed it on the lattice in
order to find its embedding number and its code (that is the set $\{\omega(o)
\}$ for all shadows).

The embedding number of two clusters on the lattice is the product
of the embedding numbers of the two clusters (modulo symmetry factors)
minus the number of embeddings where the two clusters are positioned
in such a way
that they actually form one cluster. Only the second term
is of interest because the first term is $\sim N^2$ and does therefore not
contribute to the free energy.
Instead of trying to figure out when two clusters overlap, one can work the
way backwards and determine how often a single cluster can be split into
two or more clusters.

As said above, one needs to consider only configurations
with $s \leq t$. Consequently I define the degree of an
$s$-matrix $M$
as the minimum order in $u$ of its contributions $f_{s,t}(M)$ with $t\geq s$.
The calculation of the minimum degree can be performed easily if the code
of the matrix is known.

Because all the matrices considered are finite one can show
that in each
matrix there is at least one spin which can be removed without
increasing the degree of the matrix and without disconnecting the matrix
into more than one cluster.
Turning this around, {\it all required} $(s+1)$-matrices can be constructed
by adding a spin\footnote{The conditions this spin has to fulfill will be
given elsewhere.} to $s$-matrices, and the $(s+1)$-matrices
 will have at least the degree of the $s$-matrix they were
constructed from.
This allows us to discard immediately any $s$-matrices which have
a degree higher than $l_{max}$. This in turn will prevent many
$(s+1)$-diagrams of degree $> l_{max}$ from being generated and will save
a large amount of computational effort. For some $s=s_{max}$ no matrices
with degree less than or equal to $l_{max}$ are produced when
adding another spin. At this point all connectivity
matrices have been constructed.

As was mentioned earlier I take a single cluster
apart in order to obtain disconnected diagrams. Also for this procedure
one can show that the
degree of the resulting diagram is at least the same as the degree
of the diagram one uses as starting point.

Using these two observations one is able to construct the required series
without generating unnecessary matrices or partitionings.

Despite using code balance to construct the series, it can still serve
to check the calculation:
A matrix which contributes to the order
$u^k$ of $f_{s,t}$ with
$s\geq t$ is guaranteed to be among the constructed matrices if
\be k \leq l_{max} - (s-t)(d-1).\label{kbound}\ee
This allows one to check $f_{t,s}$ against $f_{s,t}$ for orders $k$ bounded
by \eq{kbound}.

I performed the code balance test on series where some diagrams or some
partitionings were deliberately missing.
In all cases the series did not fulfill the code balance,
giving confidence in the reliability of this test.
But, as obvious from \eq{kbound}, the highest $(d-1)$ orders of a
series cannot be checked.

\section{\label{potts}}
As an example of another discrete model let me consider the 3-state Potts
model on
a $d$-dimensional lattice with $N$ sites. The action reads:

\be
 S_{h} = -\kappa \sum_{<xy>} \delta(p(x),p(y))
    - h\left(\sum_{x} \delta(p(x),1) -\frac{1}{3}N\right)
\ee
with $p(x)\in \{1,2,3\}$.

Again one starts with a ground state where all spins are aligned with the
external magnetic field. Using $u=\exp(-\kappa)$, $y=\exp(-h)$,
 the free energy for $N$ spins can be rewritten as

\be f = \frac{2}{3}N\ln y + \frac{qN}{2}\ln u
 - \sum P_{l,r,{\scriptscriptstyle N}}y^{l+r} u^t,\ee
where the $P_{l,r,{\scriptscriptstyle N}}$ are configurations with
$l$ left-pointing ($p(x)=2$)
spins and $r$ right-pointing ($p(x)=3$) spins.

The expansion is done similarly to the Ising model.
One only has to insert another step which selects a subset of the flipped
spins to be left-pointing. The shadow method can be used as well, but
instead of noting for each shadow the number of flipped neighbours, one
has to note $\omega_l$, the
number of left-pointing and $\omega_r$, the number of right-pointing
neighbours.

The minimum degree for a given $s$-matrix can be achieved by
aligning all flipped spins, for example letting them point to the left.
In this case the degree is equal to the one for the
Ising model\footnote{Up to a factor which comes from the different
definitions of $u$.}.
Therefore the set of matrices is identical to the one for the Ising model.
As in the Ising model,
the series can be checked by code balance.

\section{}
The number of matrices contributing for a given degree of the series
are given in \tab{timing}.
The time needed on a personal computer doing about 12~MIPS
is given in the last row of the table. Memory requirements were below 3Mbyte.
For the Potts model the time needed was 300 sec for $l_{max}=48$ and
6775 sec for $l_{max}=56$. This is about a factor 6 more than for the
corresponding Ising model.

Although the time needed for the d=3 Ising model
is of the order of a day,
a further extension of the series seems rather hard with this method.
Even if one assumes a 1000 MIPS machine, I would expect
1 day of CPU time to go up to $u^{36}$ and a year for $u^{40}$.

The expansion gives the free energy and all susceptibilities at
the same time, but I shall give only
the magnetisation
\be  v = \left.\langle \sigma(0) \rangle\right|_{h=0}\ee
 for the $d=3,4$ Ising model and the $d=3$ Potts model
(tables~\ref{to},\ref{tt},\ref{tq}) in this report.
Further results and an analysis of the
series will be published
elsewhere~\cite{voh} or are available by email from the author.
\\*[1cm]

{\large{\bf Acknowledgements}}\\
I would like to thank Peter Weisz for helpful discussions.
\\*[1cm]

{\large{\bf References}}

\clearpage
\begin{table}{\center
\begin{tabular}{||r|rrrr|rr||}
\hline
&\multicolumn{4}{c|}{d = 3}&\multicolumn{2}{c|}{d = 4}\\
\hline
$n\setminus l_{max}$ &  20   &   24  &  28    &   32 & 32   & 48 \\
\hline
  1 &    1  &    1  &     1  &      1  &   1 &      1 \\
  2 &    2  &    2  &     2  &      2  &   2 &      2 \\
  3 &    5  &    5  &     5  &      5  &   5 &      5 \\
  4 &   25  &   25  &    25  &     25  &  26 &     26 \\
  5 &  114  &  117  &   117  &    117  & 134 &    134 \\
  6 &  136  &  817  &   823  &    823  & 201 &   1071 \\
  7 &   11  & 1994  &  6332  &   6343  &   1 &  10010 \\
  8 &    2  &  689  & 26319  &  56200  &   1 & 113636 \\
  9 &       &   67  & 20797  & 316462  &     & 442220 \\
 10 &       &   13  &  5885  & 489566  &     &  78367 \\
 11 &       &       &   702  & 241691  &     &   2424 \\
 12 &       &       &    98  &  65554  &     &     48 \\
 13 &       &       &     2  &  10431  &     &        \\
 14 &       &       &        &   1650  &     &        \\
 15 &       &       &        &     57  &     &        \\
 16 &       &       &        &      7  &     &        \\
\hline
t[sec]& 2.0   &   43  &   1360&  91000   &2.5& 50000  \\
\hline\end{tabular} \\}

\caption{\it Number of connected diagrams for given dimension, maximum degree
 and number
of spins. Diagrams with embedding number zero are not included.\label{ndiag}
The last row gives the CPU time ($\approx$ wall-clock time) needed.
Note that the series
 with a smaller $l_{max}$ for a given dimension
 are not a prerequisite for the series with the largest $l_{max}$,
 but are listed for demonstration purposes.\label{timing}}
\end{table}

\begin{table}
\begin{centering}
\begin{tabular}{||rr|rr|rr|rr||}
\hline
 i &$v_i$ & i &$v_i$ & i &$v_i$ & i &$v_i$ \\
\hline
 0 & 1 & 9 & -792 & 17 & -9205800 & 25 & -144655483440 \\
 1 & 0 & 10 & 2148 & 18 & 30371124 & 26 & 488092130664 \\
 2 & 0 & 11 & -7716 & 19 & -101585544 & 27 & -1650000819068 \\
 3 & -2 & 12 & 23262 & 20 & 338095596 & 28 & 5583090702798 \\
 4 & 0 & 13 & -79512 & 21 & -1133491188 & 29 & -18918470423736 \\
 5 & -12 & 14 & 252054 & 22 & 3794908752 & 30 & 64167341172984 \\
 6 & 14 & 15 & -846628 & 23 & -12758932158 & 31 & -217893807812346 \\
 7 & -90 & 16 & 2753520 & 24 & 42903505030 & 32 & 740578734923544 \\
 8 & 192 &    &         &    &             &    & \\
\hline
\end{tabular}
\\ \end{centering}
\caption{\it Magnetisation $v=\sum_i v_iu^i$ for the $d=3$~Ising model}
\label{to}
\end{table}

\begin{table}
\begin{centering}
\begin{tabular}{||rr|rr|rr|rr||}
\hline
 i &$v_i$ & i &$v_i$ & i &$v_i$ & i &$v_i$ \\
\hline
 0 & 1 & 13 & -1824 & 25 & 129304000 & 37 & 15255028712400 \\
 1 & 0 & 14 & 6672 & 26 & 107955904 & 38 & -45168481179848 \\
 2 & 0 & 15 & -9216 & 27 & -1194988848 & 39 & 37715130731968 \\
 3 & 0 & 16 & -15522 & 28 & 2988132104 & 40 & 189905139915462 \\
 4 & -2 & 17 & 103920 & 29 & -1295881792 & 41 & -854318113944656 \\
 5 & 0 & 18 & -219240 & 30 & -16000351200 & 42 & 1385743813707512 \\
 6 & 0 & 19 & -6640 & 31 & 58541360096 & 43 & 1495234729403168 \\
 7 & -16 & 20 & 1433114 & 32 & -74808889446 & 44 & -14211533765551580 \\
 8 & 18 & 21 & -4364368 & 33 & -161492842096 & 45 & 34480173277650448 \\
 9 & 0 & 22 & 4015104 & 34 & 1010237004872 & 46 & -10925068439033224 \\
 10 & -168 & 23 & 16249856 & 35 & -2065384405984 & 47 & -202854191914872688 \\
 11 & 384 & 24 & -76650222 & 36 & -389570701738 & 48 & 714978709717419746 \\
 12 & -266 &   &           &    &               &    & \\
\hline
\end{tabular}
\\ \end{centering}
\caption{\it Magnetisation $v=\sum_i v_iu^i$ for the $d=4$~Ising model}
\label{tt}
\end{table}

\begin{table}
\begin{centering}
\begin{tabular}{||rr|rr|rr|rr||}
\hline
 i &$v_i$ & i &$v_i$ & i &$v_i$ & i &$v_i$ \\
\hline
   0 &  2/3 &  15 & -180 &  29 &   2153016 &  43 &    -27015116172 \\
   1 &    0 &  16 &  318 &  30 &   -792218 &  44 &    -31628035554 \\
   2 &    0 &  17 &  432 &  31 &  -8867580 &  45 &  98446988808 \\
   3 &    0 &  18 & -1320 &  32 &    935124 &  46 &    143499949662 \\
   4 &    0 &  19 & -1992 &  33 &  34889512 &  47 &   -353956661016 \\
   5 &    0 &  20 & 2760 &  34 &        63834 &  48 &   -645558433882 \\
   6 &   -2 &  21 & 9368 &  35 &   -130265472 &  49 &   1271653865928 \\
   7 &    0 &  22 & -14460 &  36 &  -39322372 &  50 &   2769960888510 \\
   8 &    0 &  23 & -35280 &  37 &  507892056 &  51 &  -4398383060152 \\
   9 &    0 &  24 &  36680 &  38 &  239776590 &  52 & -11845923653964 \\
  10 &  -12 &  25 & 134568 &  39 & -1940344524 &  53 &  14827768507104 \\
  11 &  -12 &  26 & -108516 &  40 & -1297972266 &  54 &  49906450655776 \\
  12 &   28 &  27 & -609692 &  41 & 7120754760 &  55 & -48635959747320 \\
  13 &    0 &  28 & 370500 &  42 & 6805163432 &  56 & -206168758520142 \\
  14 &    -90 &   &        &     &            &     &                  \\
\hline
\end{tabular}
\\ \end{centering}
\caption{\it Magnetisation $v=\sum_i v_iu^i$ for the $d=3$~Potts model}
\label{tq}
\end{table}

\end{document}